\documentclass[a4paper,12pt]{article}
\pdfoutput=1 

\usepackage[hmargin=.7in,vmargin=1.1in]{geometry}
\usepackage{indentfirst}
\usepackage[mathcal]{euscript}
\usepackage{amsfonts}
\usepackage{mathrsfs}
\usepackage{amsmath}
\usepackage{amssymb}
\usepackage{authblk}
\usepackage{cite}
\usepackage{xcolor}
\usepackage{mathtools}
\usepackage{tensor}
\usepackage{physics}
\usepackage{graphicx}
\usepackage{bm}
\usepackage{upgreek}
\usepackage{braket}
\usepackage{color,soul}
\usepackage{csquotes}
\usepackage{caption}
\usepackage{subcaption}
\usepackage{booktabs}
\usepackage[bookmarksnumbered=true,bookmarksopen=true]{hyperref}
\usepackage{orcidlink}
\usepackage{verbatim}

\hypersetup{
            colorlinks,
            linkcolor=[rgb]{0,0.3,0.6}, 
            citecolor=[rgb]{0,0.3,0.6}, 
            urlcolor=[rgb]{0,0.3,0.6}
           }

\def\be{\begin{eqnarray}}
\def\ee{\end{eqnarray}}

\begin{document}

\title{\Large\textbf{
Radial oscillations of quark stars in light of current astrophysical constraints: A comparative study}}

\author[a]{
Grigoris Panotopoulos  \orcidlink{0000-0002-7647-4072}
\thanks{e-mail: 
\href{mailto:grigorios.panotopoulos@ufrontera.cl}
{\nolinkurl{grigorios.panotopoulos@ufrontera.cl}}}
}

\author[b, c
]{
\'Angel Rinc\'on {\orcidlink{0000-0001-8069-9162}} \thanks{e-mail: 
\href{mailto:angel.rincon@physics.slu.cz}
{\nolinkurl{angel.rincon@physics.slu.cz}}}
}

\author[d
]{
Ilidio Lopes  \orcidlink{0000-0002-5011-9195}
\thanks{e-mail: 
\href{mailto:ilidio.lopes@tecnico.ulisboa.pt}
{\nolinkurl{ilidio.lopes@tecnico.ulisboa.pt}}}
}

\affil[a]{\normalsize{\em Departamento de Ciencias F{\'i}sicas, Universidad de La Frontera, Casilla 54-D, 4811186 Temuco, Chile.}}

\affil[b]{\normalsize{
\em Departamento de F{\'i}sica, Universidad del B{\'i}o-B{\'i}o,
Casilla 5-C, Concepci{\'o}n, Chile.}
}

\affil[c]{\normalsize{
\em Research Centre for Theoretical Physics and Astrophysics, Institute of Physics, Silesian University in Opava, Bezručovo náměstí 13, CZ-74601 Opava, Czech Republic.}
}


\affil[d
]{\normalsize{\em Centro de Astrof{\'i}sica e Gravita{\c c}{\~a}o, Departamento de F{\'i}sica, Instituto Superior T{\'e}cnico-IST, Universidade de Lisboa-UL, Av. Rovisco Pais, 1049-001 Lisboa, Portugal.}}

\date{ }

\maketitle

\begin{abstract}

\smallskip\noindent	

We investigate the structural and oscillatory properties of isotropic strange quark stars within General Relativity, focusing on three physically motivated equations of state: the colour-flavour-locked (CFL) phase, an interacting quark matter model, and a linear (causal) equation of state. By numerically solving the Tolman–Oppenheimer–Volkoff and radial perturbation equations, we construct equilibrium stellar sequences and compute oscillation spectra across three representative masses (0.77, 1.4, and 2.0~$M_\odot$). Our analysis is focused on two diagnostics: (i) mass–radius profiles and (ii) radial mode eigenfrequencies with large frequency separations. We compare theoretical predictions against multi-messenger constraints from NICER X-ray timing of key pulsars, the massive pulsars at two solar masses, and the low-mass compact object in HESS~J1731$-$347. All three equations of state yield maximum masses exceeding 2~$M_\odot$ with canonical-mass radii of $(10–12)~km$, satisfying current observational bounds. Fundamental mode frequencies span 4–7~kHz, with asymptotic large separations differing among the models. These elevated frequencies lie within the detection band of current and next-generation gravitational-wave observatories, offering potential asteroseismic signatures for distinguishing strange quark stars from hadronic neutron stars in post-merger emission. Our results demonstrate that self-bound quark matter naturally accommodates the sub-solar mass configuration of HESS~J1731$-$347, reinforcing the viability of strange quark star interpretations.
\end{abstract}


\tableofcontents

\newpage

\section{Introduction}
\label{sec:intr}


Neutron stars are the ultra-compact endpoints of stellar evolution for progenitor stars with initial masses between approximately 8 and 25\,$M_\odot$. Following a core-collapse supernova, the remaining core, typically of mass (1.1 - 2)\,$M_\odot$, collapses into a sphere of only $\sim$ 10 - 14\,km radius. This implies central densities several times above the nuclear saturation density $\rho_0 \approx 2.8 \times 10^{14}$\,g\,cm$^{-3}$ (or roughly 0.16 nucleons\,fm$^{-3}$) 
\cite{Shapiro:1983du,2000csnp.conf.....G,Haensel2007,2012ARNPS..62..485L,2025RvMP...97d5007C}. Under these extreme conditions, the composition transcends ordinary nuclear matter: the stellar core is a degenerate, strongly interacting, neutron-rich fluid that may contain hyperons, Bose condensates, deconfined quark matter, or other exotic phases whose existence and properties are dictated by the largely unknown \emph{equation of state} (EoS) of cold matter beyond nuclear saturation density
\cite{2004Sci...304..536L,2016PhR...621..127L,2021ARNPS..71..433L}.

\smallskip

The equation of state at supranuclear density constitutes one of the central open questions in nuclear astrophysics and quantum chromodynamics (QCD). It governs the global macroscopic parameters of the star, the mass-radius relation, moment of inertia, maximum mass, and tidal deformability, whilst simultaneously controlling the microphysics of the interior, including the speed of sound, symmetry energy, superfluid gaps, and possible first- or second-order phase transitions. Constraining the EoS is therefore essential not only for understanding neutron-star structure, but also for testing fundamental physics under conditions unattainable in terrestrial laboratories: strong-field gravity, the strong interaction at high baryon density, and the interplay between general relativity and nuclear many-body theory \cite{2017RvMP...89a5007O,2016PhR...621..127L,2021PrPNP.12003879B,2020PhRvD.101f3025S,2024JCAP...05..130R,2025arXiv250903517G,2025arXiv251207977G}.

\smallskip

Oscillation modes of neutron stars, collectively referred to as \emph{neutron-star asteroseismology}, provide a uniquely powerful and essentially direct probe of the interior physics, complementary to electromagnetic radius measurements (e.g.\ NICER, Athena) and gravitational-wave constraints on tidal deformability (LIGO--Virgo--KAGRA). Just as Helioseismology has revolutionized our knowledge of the solar interior \cite{1993ApJ...408..347T}, neutron-star seismology can, in principle, map the internal composition, stratification, elastic properties, superfluid state, and possible phase transitions by identifying distinct families of oscillation modes and relating their frequencies and damping rates to the underlying EoS and microphysics \cite{Andersson:1997rn}.

\smallskip

The principal mode families in relativistic stars \cite{1998MNRAS.299.1059A,1999LRRel...2....2K,1983ApJ...268..837M,2011PhRvD..83b4014S} include:
(i) p-modes (pressure modes), primarily acoustic waves restored by pressure gradients;
(ii) g-modes (gravity modes), restored by buoyancy in compositionally or thermally stratified regions;
(iii) f-modes (fundamental pressure modes), which are particularly sensitive to the stellar mean density;
(iv) s-modes and t-modes (shear and torsional modes of the elastic crust);
(v) i-modes (interface modes at sharp phase transitions, e.g.\ hadron--quark);
and (vi) w-modes (spacetime modes, pure gravitational-wave excitations of the curved background).
Among these, radial oscillation modes are of particular interest because they are directly linked to the stellar stability criterion $\partial M/\partial \rho_c > 0$, with the fundamental radial mode frequency vanishing precisely at the maximum-mass configuration \cite{moustakidis2017_stability}.

\smallskip

In the context of General Relativity several works have investigated compact stars and their radial oscillations using different equations-of-state (EoSs) and approaches. For instance:
(i) In \cite{DiClemente:2020szl}, the authors developed a numerical algorithm to solve the Sturm-Liouville differential equation governing the stationary radial oscillations of non-rotating compact stars.
(ii) In \cite{Bora:2020cly}, the radial oscillations of non-rotating strange stars and their characteristic echo frequencies were studied considering three different EoSs: the MIT Bag model EoS, a linear EoS, and a polytropic EoS.
(iii) Additional studies employing the CFL EoS can be found in \cite{Arbanil:2016wud,Zhang:2021iah,Lopes:2019psm,Rocha:2019cuq}, see also \cite{Rather1,Rather2,Rather3,Routaray,Sepulveda}.

\smallskip

In the literature there is a vast amount of articles in which the authors have studied interior solutions and structural properties of relativistic stars in different context. To be more precise, thanks to the work of \cite{Ovalle:2017fgl}, an elegant method has been developed to generate anisotropic solutions from a known isotropic seed solution. The Minimal Geometric Deformation approach, originally introduced in \cite{Ovalle:2007bn} within the brane-world context, has since become a powerful and convenient tool for studying self-gravitating systems such as relativistic stars and black holes \cite{Estrada:2018zbh,Morales:2018urp,Estrada:2018vrl,Ovalle:2018umz,Misyura:2024fho,Tello-Ortiz:2023yxz,Gabbanelli:2018bhs}. Its later extensions and applications \cite{Ovalle:2018gic,Fernandes-Silva:2019fez} further proved its versatility and effectiveness.
More recently, the complexity factor introduced in \cite{Herrera:2018bww} provided a way to quantify the structural complexity of self-gravitating systems. Emerging from the orthogonal splitting of the Riemann tensor, it vanishes for isotropic or homogeneous energy-density configurations, but may also vanish when anisotropy and density inhomogeneity compensate each other. Numerous works have applied this approach to anisotropic stellar models \cite{Abbas:2018idr,Sharif:2018pgq,Abbas:2018cha,Nazar:2019cma,Arias:2021gth,Rincon:2023zlp,Rincon:2023ens}.
Another method for obtaining exact interior solutions of relativistic stars is based on the Karmarkar condition \cite{Karmarkar1948}. By specifying one metric potential, the condition determines the other, allowing the energy density and pressures to be computed from the field equations without assuming an equation of state. This approach has been widely used to construct interior solutions in gravitational theories \cite{Maurya:2015qfm,Maurya:2016zdw,Bhar:2017ynp,Tello-Ortiz:2019gcl,Baskey:2020guh}.

\smallskip

The motivation for intensive study of neutron-star oscillation modes has grown dramatically over the past two decades. This growth has been driven by the discovery of quasi-periodic oscillations (QPOs) in giant flares from magnetars \cite{Israel:2005av,Strohmayer:2007qj}, the advent of gravitational-wave astronomy marked by the landmark detection GW170817 \cite{LIGOScientific:2017vwq}, and the promise of future high-precision X-ray timing missions and third-generation gravitational-wave detectors that may directly detect normal modes. These developments collectively illustrate that the study of neutron-star oscillation modes has evolved from a theoretical curiosity in the 1990s to one of the most promising and observationally accessible pathways toward deciphering the supranuclear equation of state and the exotic physics of matter at extreme densities in the multi-messenger era.

\smallskip

A particularly intriguing possibility is that compact stars may harbour deconfined quark matter in their cores, or may even be composed entirely of self-bound strange quark matter, so-called strange quark stars. The hypothesis that strange quark matter could be the true ground state of strongly interacting matter dates back to seminal works by Bodmer, Witten, and Terazawa \cite{Bodmer:1971we,Witten:1984rs,Terazawa:1989iw}. At asymptotically high densities, QCD predicts that quark matter enters a colour superconducting phase, with the color-flavor-locked (CFL) state being the most symmetric pairing pattern \cite{Rajagopal:2000ff}. The CFL phase, in which all three quark flavors and colors participate in Cooper pairing, is believed to be the ground state at sufficiently high baryon densities, and its properties have important implications for compact star phenomenology \cite{Lugones:2002va,VasquezFlores:2017uor}.

\smallskip

Relativistic stars provide an effective natural laboratory to test strong-field gravity. Key observables such as the mass-radius relation, compactness, and tidal deformability can be directly compared with constraints from radio pulsar timing, X-ray measurements, and gravitational-wave observations \cite{LIGOScientific:2020zkf,Miller:2021qha,Riley:2021pdl}. The importance of these tests has been reinforced by multi-messenger detections of binary neutron star mergers \cite{2018PhRvL.121i1102D}. Recent X-ray timing observations from the Neutron Star Interior Composition Explorer (NICER) have placed stringent constraints on neutron star radii: PSR~J0740+6620, the most massive precisely measured neutron star at $2.08 \pm 0.07\,M_\odot$, has an inferred radius of approximately $12.5$--$13\,\text{km}$ \cite{2024ApJ...974..295D,2024ApJ...974..294S}, whilst PSR~J0030+0451 and PSR~J0437$-$4715 provide complementary constraints at lower masses \cite{Miller:2019cac,Choudhury:2024xbk}. The discovery of the unusually light compact object in the supernova remnant HESS~J1731$-$347, with an inferred mass of approximately $0.77\,M_\odot$ \cite{Doroshenko:2022nwp}, has sparked renewed interest in strange quark star models, as such low masses are difficult to reconcile with standard hadronic equations of state but arise naturally in self-bound quark matter scenarios \cite{Rather:2023epjc}.

\smallskip

Recent studies have demonstrated that asteroseismic observables, in particular, the large frequency separation $\Delta\nu $, encode valuable information about the stellar mean density and sound speed profile, thereby serving as indirect probes of the underlying EoS \cite{Sagun:2020txh,Rather:2024jcap}. For quark stars, the characteristic oscillation frequencies lie in the kilohertz range, within the detection band of current and next-generation gravitational-wave observatories \cite{LIGOScientific:2017vwq}. The elevated fundamental mode frequencies typical of quark matter compositions, arising from their stiffer sound speed profiles at high densities, could potentially serve as distinguishing signatures for differentiating quark stars from their hadronic counterparts in future gravitational-wave detections.

\smallskip

In light of current astrophysical constraints, we consider here three analytic EoS for quark matter that predict mass-to-radius relationships in agreement with several observational data. Next, we compute and compare the spectra of their radial oscillation modes. To the best of our knowledge, this kind of comparative study is performed for the first time in the present work. The goal of this article is to conduct a systematic comparative study of compact stars composed of isotropic quark matter, focusing on the impact of three distinct realistic equations-of-state, namely the color-flavor-locked (CFL) phase, an interacting quark matter model, and a linear (causal) EoS, on stellar structure and radial oscillation properties. We seek to determine whether these different quark matter descriptions yield observationally distinguishable signatures in the mass--radius relation and oscillation spectra, and to assess their compatibility with current multi-messenger astrophysical constraints.

\smallskip

The structure of our paper is as follows. In Section~\ref{GR}, we review the theoretical framework for stellar equilibrium and radial oscillations in four-dimensional general relativity, presenting the Tolman--Oppenheimer--Volkoff (TOV) equations for hydrostatic equilibrium in Section~\ref{stellareq} alongside the boundary conditions necessary for stellar equilibrium, and the Sturm--Liouville eigenvalue problem governing radial perturbations. In Section~\ref{sec:pert}, we introduce the three equations of state under consideration, the CFL, interacting, and linear EoS, provide the physical motivation for their selection, and highlight their relevance to realistic compact star modeling. Section~\ref{sec:meth} contains the numerical results, including stellar mass-radius profiles, and the radial oscillation eigenfrequencies and large frequency separations for representative stellar masses. Finally, in Section~\ref{sec:disc}, we analyse the implications of these results and summarise our conclusions, with emphasis on future directions such as tidal deformability studies and rotating stellar configurations.

\section{Stellar equilibrium and radial oscillations}\label{GR}


In what follows we shall review the relevant equations useful for describing relativistic stars in the context of four-dimensional General Relativity without a cosmological constant.

\subsection{Interior solutions of fluid spheres}\label{stellareq}

In the present work we shall {\bf be} studying electrically neutral and non-rotating stars made of a single fluid component. Moreover, we shall assume that the stars are made of isotropic matter, and so´the matter content is described by the following stress-energy tensor
\begin{align}
 T^{\mu}_{\nu} \equiv \text{diag} \{ -\rho(r), P(r), P(r), P(r) \}.
\end{align}
where the labels have the following meaning: 
i) $\rho(r)$ is the energy density of the fluid,  
ii) $P(r)$ is the pressure of the fluid, both in radial and tangential directions.
We note in passing that for the sake of simplicity, we have chosen to neglect the viscosity term.
Next, we shall assume static, spherically symmetric astronomical objects, in Schwarzschild-like coordinates, where $x^0=t; \, x^1=r; \, x^2=\theta; \, x^3=\phi$. The metric tensor is then given by the usual expression
\begin{equation}
\mathrm{d}s^2 = -e^{\nu} \mathrm{d}t^2 + e^{\lambda} \mathrm{d}r^2 +
r^2 \mathrm{d}\Omega^2,
\label{metric}
\end{equation}
where $d\Omega^2\equiv \left( d\theta^2 + \sin^2\theta d\phi^2 \right)$ is the line element of the two-dimensional unit sphere, while the metric potentials, $\nu(r)$ and $\lambda(r)$, depend on the radial coordinate only.
Using Einstein's field equations (with the cosmological constant set to zero), the components of the energy-momentum tensor can be expressed (in local Minkowski coordinates and in terms of the metric potentials) as follows:
\begin{eqnarray}
\rho &=& -\frac{1}{8\pi}\left[-\frac{1}{r^2}+e^{-\lambda}
\left(\frac{1}{r^2}-\frac{\lambda'}{r} \right)\right],
\label{fieq00}
\\
P &=& -\frac{1}{8\pi}\left[\frac{1}{r^2} - e^{-\lambda}
\left(\frac{1}{r^2}+\frac{\nu'}{r}\right)\right],
\label{fieq11}
\\
P &=& \frac{1}{32\pi}e^{-\lambda}
\left(2\nu''+\nu'^2 -
\lambda'\nu' + 2\frac{\nu' - \lambda'}{r}\right),
\label{fieq2233}
\end{eqnarray}
where a prime represents differentiation with respect to the radial coordinate $r$. 

At this stage, we derive the anisotropic hydrostatic equilibrium equation, namely the generalized Tolman–Oppenheimer–Volkoff (TOV) equation for anisotropic stars.
Combining Eqs.~\eqref{fieq00}, \eqref{fieq11}, and \eqref{fieq2233}, we obtain:
\begin{equation} \label{Prp}
P' = -\frac{1}{2}\nu'\left( \rho + P \right) .  
\end{equation}
Even more, we can eliminate the $\nu'$-dependence in equation (\ref{Prp}) taking advantage of the following equation
\begin{equation}
\frac{1}{2}\nu' =  \frac{m + 4 \pi P r^3}{r \left(r - 2m\right)}.
\label{nuprii}
\end{equation}
Thus, substituting the \eqref{nuprii} into \eqref{Prp}, we find the anisotropic TOV equation, namely
\begin{equation} \label{ntov}
P'=-\frac{m + 4 \pi P r^3}{r \left(r - 2m\right)} \; \left( \rho + P \right) .
\end{equation}
The system of stellar structure equations is completed by including the differential equation for the mass function $m(r)$, namely:
\begin{align}
    m' &= 4 \pi r^2 \rho, \label{eq_mass}
\end{align}
which as usual is defined by
\begin{equation}
1-e^{-\lambda}=\frac{2m}{r}.
\label{rieman}
\end{equation}
Note that the TOV equation can be rewritten in a form that explicitly separates the contributions from the different forces acting within the star. In particular, it is convenient to identify two distinct terms:
  i) gravitational, $F_g$, and 
 ii) hydrostatic, $F_r$, 
defined according to the expressions
\begin{subequations}
\begin{align}
F_g &= -\left( \frac{m + 4 \pi P r^3}{r \left(r - 2m\right)} \right)\left( \rho + P \right) ,
\\
F_r &= -P' .
\end{align}
\end{subequations}
Combining the last definition for two three forces and plug them into the equation \eqref{ntov}, we can obtain the expression:
\begin{equation}
F_g + F_r = 0,\label{Frp}
\end{equation}	
as was mentioned in \cite{Prasad:2021eju}.
Finally, the exterior vacuum solution is given by the Schwarzschild geometry, i.e.,
\begin{equation}
\mathrm{d}s^2= -\left(1-\frac{2M}{r}\right) \mathrm{d}t^2 + \left(1-\frac{2M}{r}\right)^{-1} \mathrm{d}r^2 +
r^2  \mathrm{d}\Omega^2.
\label{Vaidya}
\end{equation}
with $M$ being the stellar mass. Thus, the system of equations is closed imposing appropriate matching conditions at the surface $r=r_\Sigma \equiv  R$ for the Schwarzschild spacetime, where $R$ is the stellar radius.
With the help of the first and the second fundamental forms across that surface we have
\begin{subequations}
\begin{eqnarray}
e^{\nu(r)} \Bigl|_{r=R} &=& 1-\frac{2M}{R},
\label{enusigma}
\\
e^{\lambda(r)} \Bigl|_{r=R} &=& \Bigg(1-\frac{2M}{R}\Bigg)^{-1},
\label{elambdasigma}
\\
P(r) \Bigl|_{r=R} &=& 0.
\label{PQ}
\end{eqnarray}
\end{subequations}
The last three equations are the necessary (and sufficient) conditions for a smooth matching of the two metrics (\ref{metric}) and (\ref{Vaidya}) on the surface $r=R$.

\subsection{Equations for perturbations of pulsating stars} \label{radial}

Considering a spherically symmetric system with only radial motion, Einstein's field equations can be used to compute the radial oscillation properties for a static equilibrium structure \cite{1966ApJ...145..505B}.
Considering the radial displacement $\Delta r$ with the pressure perturbation as $\Delta P$, the small perturbation of the equations governing the dimensionless quantities $\xi$ = $\Delta r/r$ and $\eta$ = $\Delta P/P$ are defined as \cite{1977ApJ...217..799C, Gondek:1997fd} 
\begin{equation}\label{ksi}
     \xi'(r) = -\frac{1}{r} \Biggl( 3\xi + \frac{\eta}{\Gamma} \Biggr) - \frac{P'(r)}{P+\rho} \xi(r),
\end{equation}
\begin{equation}\label{eta}
    \begin{split}
          \eta'(r) = &\xi \Biggl[ \omega^{2} r (1+\rho/P) e^{\lambda - \nu } -\frac{4P'(r)}{P} -8\pi (P+\rho) re^{\lambda} 
     +  \frac{r(P'(r))^{2}}{P(P+\rho)}\Biggr] + 
     \\
     &\eta \Biggl[ -\frac{\rho P'(r)}{P(P+\rho)} -4\pi (P+\rho) re^{\lambda}\Biggr] ,
    \end{split}
\end{equation}
where $\omega$ is the frequency oscillation mode, while $\Gamma$ is the adiabatic relativistic index
\begin{equation}
     \Gamma = \Biggl(1+\frac{\rho}{P} \Biggr) c_s^{2},
 \end{equation}
with $c_s^{2}$ as the speed of sound squared given by
\begin{equation}\label{cs}
     c_s^{2} = \Biggl(\frac{dP}{d\rho}\Biggr).
\end{equation}
These two coupled differential equations, Eqs.~\eqref{ksi} and \eqref{eta}, are supplemented with two boundary conditions: one at the center, where $r$ = 0, and another at the surface, where $r$ = $R$. The boundary condition at the center requires that
\begin{equation}
    \eta = -3\Gamma \xi 
\end{equation}
must be satisfied. The equation Eq.~\eqref{eta} must be finite at the surface and hence
\begin{equation}
    \eta = \xi \Biggl[ -4 + \Bigg(1-\frac{2M}{R}\Bigg)^{-1} \Biggl( -\frac{M}{R} -\frac{\omega^{2} R^{3}}{M}\Biggr)\Biggr]
\end{equation}
must be satisfied where $M$ and $R$ correspond to the mass and radius of the star, respectively. The frequencies are computed by
\begin{equation}
\nu = \frac{\omega}{2\pi} = \frac{s \: \omega_0}{2\pi} ~~(\text{kHz}),
\end{equation}
where $s$ is a dimensionless number, while $\omega_0 \equiv \sqrt{M/R^3}$.

We use the shooting method analysis, where one starts the integration for a trial value of $\omega^2$ and a given set of initial values that satisfy the boundary condition at the center. We integrate towards the surface, and the discrete values of $\omega^2$ for which the boundary conditions are satisfied correspond to the eigenfrequencies of the
radial perturbations.

These equations represent the Sturm-Liouville eigenvalue equations for $\omega$. The solutions provide the discrete eigenvalues $\omega_n^{2}$ and can be ordered as 
\begin{equation*}
\omega_0 ^{2} < \omega_1 ^{2} <... <\omega_n ^{2}, 
\end{equation*}
where $n$ is the number of nodes for a star of a given mass and radius. Finally, once the spectrum is known, the so called large frequency separation may be computed
\begin{equation}
\Delta \nu_n = \nu_{n+1} - \nu_n, \; \; \; n=0,1,2,3,...
\end{equation}
or in other words the difference between consecutive modes, which is widely used in Asteroseismology, and which can be shown to relate to stellar mass and radius \cite{tassoul, miglio, ilidio}.

\section{Matter content: Equation of state}
\label{sec:pert}

The equation of state (EoS) plays a crucial role in compact star physics by describing matter at extreme densities and determining stellar mass and radius, and stability of the configuration. More precisely, the EoS is a non-trivial relationship connecting thermodynamic variables that specify the state of a physical system. Broadly speaking, the EoS can be mathematically expressed as an expansion of pressure in powers of density, where the coefficients in the series encode deviations from the simplest scenario and can be derived from the underlying elementary interactions.
Consequently, the EoS encapsulates essential dynamical information, enabling a suitable choice to link measurable macroscopic quantities with the forces acting between the constituents of the system at microscopic level.

Strange quark stars are based on the seminal works from the 70s and 80s \cite{Itoh:1970uw,Bodmer:1971we,Witten:1984rs,Terazawa:1989iw}, where it was proposed that quark matter is by assumption absolutely stable, and as such it may be the true ground state of Quantum Chromodynamics (QCD). According to this idea, up, down and strange quarks in weak equilibrium, in the stellar interior become effectively massless as compared to the associated chemical potential at very large densities, forming Cooper pairs with a common Fermi momentum. Since those pairs are electrically neutral, electrons cannot be present in this superfluid ground state \cite{Rajagopal:2000ff}, dubbed color-flavor-locked (CFL) phase.

\smallskip\noindent

Now, let us take three particular EoS to be used in what follows. All of them are equations-of-state that go beyond the simplest MIT bag model "radiation plus constant" taking into account interactions. Therefore, they are more realistic, while at the same time they are analytic and capable of supporting configurations in agreement with current astrophysical constraints, see the first figure below.

\subsection{Quark matter: CFL EoS}

For CFL strange stars, the equation of state of quark matter may be obtained within the framework of the MIT bag model \cite{Chodos:1974pn,Chodos:1974je,Johnson1975}, although now due to QCD superconductivity effects the linear EoS of the simplest version receives corrections of order $(\Delta/\mu)^2$, which is around a few percent for typical values of the color superconducting energy gap ($\Delta \sim 0-150$ MeV) and the baryon chemical potential ($\mu \sim 300-400$ MeV). From all the possible viable models, we shall adopt here the model CFL 3 \cite{VasquezFlores:2017uor},  for which the numerical values of the 3 parameters are as follows:
\begin{equation}
B = 60 \ \frac{\text{MeV}}{\text{fm}^3}, \; \; \; \; \Delta = 100 \ \text{MeV},  \; \; \; \; m_s = 0 \ \text{MeV},
\end{equation}
with $m_s$ being the mass of the s quark, and $B$ being the bag constant.
At order $\Delta^{2}$ and  $m^{2}_{s}$ the pressure and energy density can be written as \cite{2002PhRvD..66g4017L}:
\begin{equation}\label{PB}
P=\frac{3\mu^{4}}{4\pi^{2}}+\frac{9\alpha \mu^{2}}{2\pi^{2}} - B,
\end{equation}
\begin{equation}\label{energiaB}
\rho = \frac{9\mu^{4}}{4\pi^{2}}+\frac{9\alpha\mu^{2}}{2\pi^{2}} + B,
\end{equation}
where
\begin{equation}\label{alfa}
\alpha = -\frac{m^{2}_{s}}{6}+\frac{2\Delta^{2}}{3}.
\end{equation}
From the above expressions, we can obtain an analytic expression for $\rho = \rho(P)$: 
\begin{equation}\label{energiaC}
\rho = 3P+4B-\frac{9\alpha\mu^{2}}{\pi^{2}},
\end{equation}
with
\begin{equation}\label{mu2}
\mu^{2}= -3\alpha +\bigg[\frac{4}{3}\pi^{2}(B+P)+9\alpha^{2}\bigg]^{1/2},
\end{equation}

\smallskip\noindent 
to finally obtain
\begin{equation}
\rho = 3P+4B-\frac{9\alpha}{\pi^{2}} 
\Bigg[-3\alpha +\sqrt{\frac{4}{3}\pi^{2}(B+P)+9\alpha^{2}}
\Bigg].
\end{equation}


\begin{figure*}[h]
\centering
\includegraphics[width=0.80\textwidth]{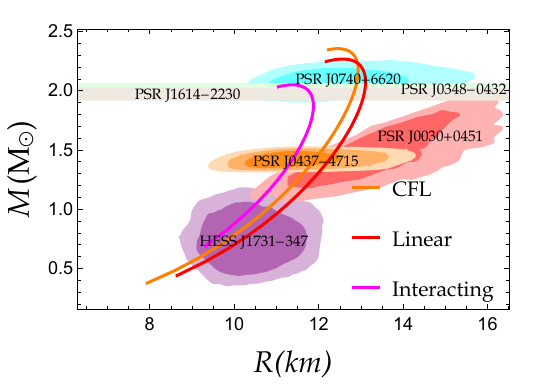}
\caption{
\textcolor{black}{
Mass-to-radius relationships for the three EoSs considered in this work.
In addition, there are four contour plots corresponding to:
a) the light HESS compact object (purple region) \cite{Doroshenko:2022nwp},
b) the pulsar J0740+6620 (cyan region) \cite{Miller:2021qha,Riley:2021pdl}, 
c) the pulsar J0030+0451 (red region), \cite{Miller:2019cac}, 
and d) the pulsar J0437-4715 (orange region) \cite{Choudhury:2024xbk}.
The intensity of the color represents 65$\%$, 90$\%$ and 99$\%$ 
confidence levels (CLs) from darker to lighter color, delineating the observationally allowed mass-radius parameter space for each compact object. We have also included the most massive pulsars at around two solar masses, namely the PSR J1614-2230 which has a known mass of $M=(1.97\pm0.04)M_{\odot}$ \cite{Demorest} as well as the PSR J0348+0432, which has a known mass of $M=(2.01\pm0.04)M_{\odot}$ \cite{Antoniadis:2013pzd}.}
}
\label{fig:MR-profile1}
\end{figure*}


\subsection{Quark matter: Interacting EoS}

We again compute numerically interior solutions of realistic spherical configurations of anisotropic matter. For quark matter we adopt the interacting equation-of-state, $P(\rho)$, given by \cite{Becerra-Vergara:2019uzm,Panotopoulos:2021cxu}
\begin{eqnarray} \label{Prad1}
&& P = \dfrac{1}{3}\left(\rho-4B\right)-\dfrac{m_{s}^{2}}{3\pi}\sqrt{\dfrac{\rho-B}{a_4}} 
+\dfrac{m_{s}^{4}}{12\pi^{2}}\left[1-\dfrac{1}{a_4}+3\ln\left(\dfrac{8\pi}{3m_{s}^{2}}\sqrt{\dfrac{\rho-B}{a_4}}\right)\right],
\end{eqnarray}
where $P$ is the pressure, and $\rho$ is the energy density of homogeneously distributed quark matter (also to $\mathcal{O}$ $(m_s^4)$ in the Bag model). For the purpose of the present analysis, following Beringer \textit{et al} \cite{ParticleDataGroup:2012pjm}, we take the strange quark mass to be $m_{s}$ to be $100 \,{\rm MeV}$, while the accepted values of the bag constant, $B$, lies within the range of \cite{FiorellaBurgio:2018dga,Blaschke:2018mqw} 
\begin{align}
57 \ \frac{\text{MeV}}{\text{fm}^3} \leq B \leq 92 \ \frac{\text{MeV}}{\text{fm}^3} .
\end{align}
The parameter $a_4$ comes from QCD corrections on the pressure of the quark-free Fermi sea, and it is directly related with the mass-radius relations of Quark Stars. Note that as far as the EoS is concerned, only the radial pressure is relevant. The tangential pressure has already defined in the Introduction as well as after eq. (2).

\begin{figure*}
	\centering
	\includegraphics[width=0.325\textwidth]{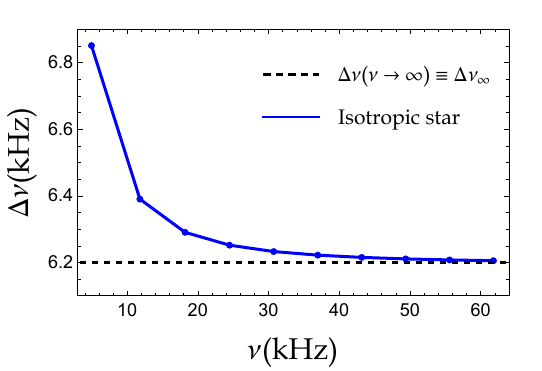}
	\includegraphics[width=0.325\textwidth]{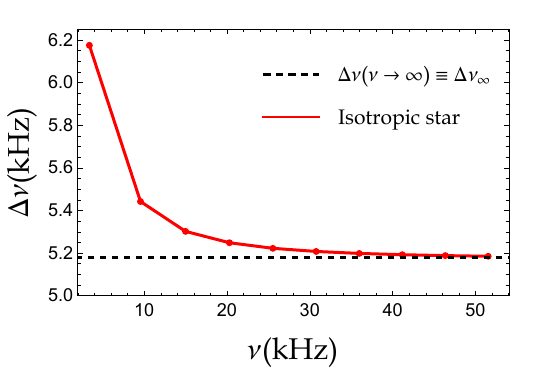}
   	\includegraphics[width=0.325\textwidth]{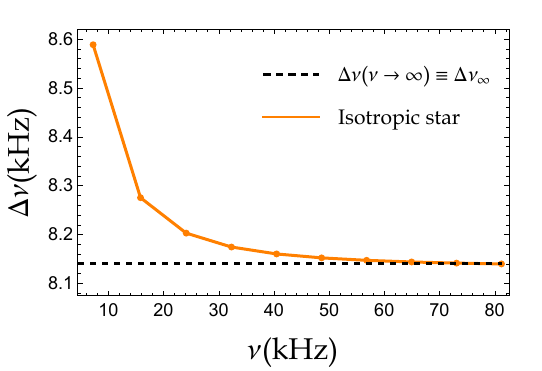}
    \\
    \includegraphics[width=0.325\textwidth]{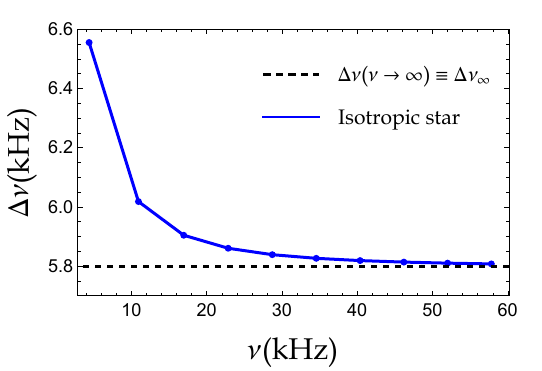}
	\includegraphics[width=0.325\textwidth]{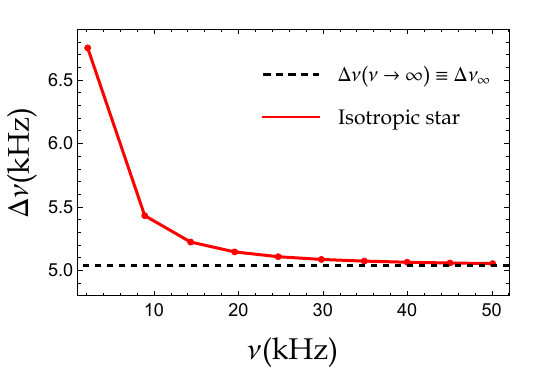}
   	\includegraphics[width=0.325\textwidth]{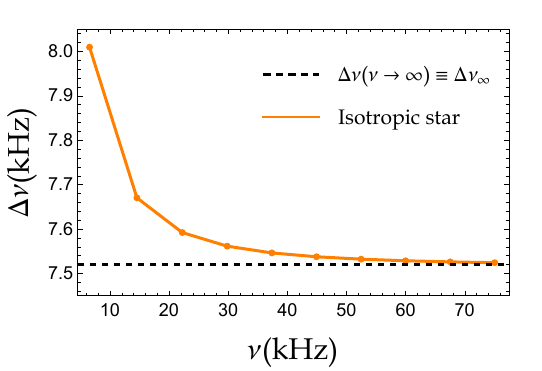}
    \\
    \includegraphics[width=0.325\textwidth]{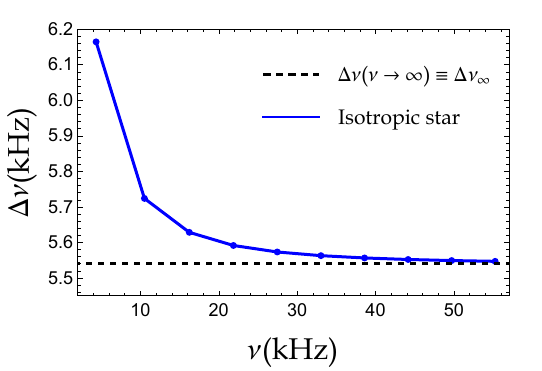}
	\includegraphics[width=0.325\textwidth]{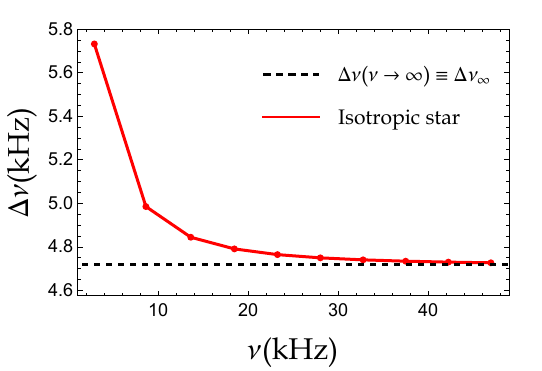}
   	\includegraphics[width=0.325\textwidth]{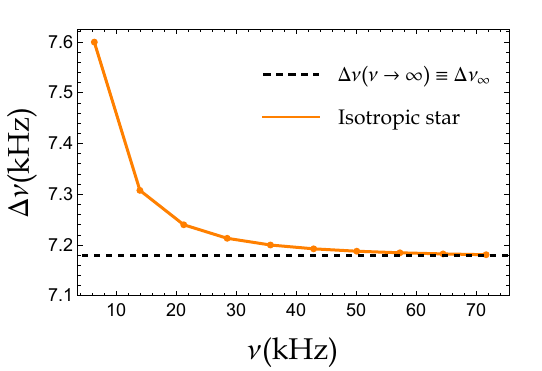}
	\caption{
		\textcolor{black}{
			{\bf{Top row:}} 
            Large frequency separations versus frequency for the CFL EoS for three different masses $M=1.4 \ M_{\odot}$ (left),  $M=2.0\ M_{\odot}$ (Middle) and  $M=0.77 \ M_{\odot}$ (right). At higher excited modes the separation tends to a constant value, $\Delta \nu_{\infty} = 6.20$ kHz (left), $\Delta \nu_{\infty} = 5.18$ kHz (middle) and finally             $\Delta \nu_{\infty} = 8.14$ kHz (right).
			{\bf{Middle panel:}} Same as before, but for the interacting EoS. This time at higher excited modes the asymptotic value of frequency separations are found to be: $\Delta \nu_{\infty} = 5.80$ kHz (left), $\Delta \nu_{\infty} = 5.04$ kHz (middle) and finally $\Delta \nu_{\infty} = 7.52$ kHz (right). 
			{\bf{Down panel:}} Same as before, but for the linear EoS. At higher excited modes the asymptotic value of frequency separations are computed to be: $\Delta \nu_{\infty} = 5.54$ kHz (left), $\Delta \nu_{\infty} = 4.72$ kHz (middle) and finally $\Delta \nu_{\infty} = 7.18$ kHz (right). 
		}
	}
	\label{fig:MR-profile2}
\end{figure*}

\subsection{Quark matter: Linear EoS}

Quark stars are hypothetical compact objects composed of deconfined quark matter rather than hadronic matter, potentially existing when extreme densities overcome quark confinement. The equation of state (EOS) relating pressure and energy density is critical for determining their mass and radius. A simple linear EOS of the form $P = k \rho$ (with $k \approx 1/3$) arises naturally in models like the MIT bag model for strange quark matter and offers significant advantages for theoretical study.
This linear relation allows analytical or semi-analytical solutions to the Tolman-Oppenheimer-Volkoff equations, yielding compact stars that can reach masses above 2 solar masses with smaller radii than conventional neutron stars. Recent multi-messenger observations-gravitational waves from mergers and precise radius measurements from NICER-tightly constrain the EOS, making the distinct predictions of a linear EOS highly testable. Additionally, a linear EOS facilitates efficient exploration of hybrid stars, rotating configurations, and the possible stability of strange quark matter as the true ground state of strongly interacting matter (Bodmer-Witten hypothesis). Thus, studying a linear EOS for quark stars provides a valuable bridge between QCD at high density and observable astrophysical phenomena.

Taking the latter into account, we will take advantage of that and we will follow the suggestion of \cite{Gondek-Rosinska:2008zmv}, which is
\begin{align}
P &=  a(\rho-\rho_0), 
\end{align}
where the numerical values are as follows
\begin{align}
a=0.324, 
\hspace{1cm}
\rho_0 = 3.0563 \times 10^{17} \ \frac{\text{kg}}{\text{m}^3}.
\end{align}

\section{Numerical Solutions}
\label{sec:meth}

\smallskip\noindent

We present numerical results obtained by integrating the relativistic stellar structure equations in full General Relativity (GR) for three distinct equations of state (EoS) describing isotropic quark matter. Specifically, the following EoS prescriptions are employed: first, a color-flavor-locked (CFL) phase utilising the CFL~3 parameterisation developed within the MIT bag model framework, which incorporates perturbative QCD corrections up to order $(\Delta/\mu)^2$~\cite{VasquezFlores:2017uor,Lugones:2002va}; second, an interacting quark matter model that accounts for density-dependent quark masses and vector-type interactions 
~\cite{Becerra-Vergara:2019uzm,Panotopoulos:2021cxu}; and third, a simple linear (or causal) EoS of the form $p = a(\rho - \rho_0)$, commonly adopted as a stiff, schematic representation of strange quark matter~\cite{Gondek-Rosinska:2008zmv}. Throughout this analysis, the pressure is assumed to be isotropic, i.e.\ $p_r = p_\perp = p$.

Static stellar configurations are constructed by numerically solving the Tolman--Oppenheimer--Volkoff (TOV) equations [Eqs.~\eqref{eq_mass} and \eqref{ntov}] subject to standard boundary conditions: vanishing enclosed mass at the center of the star, $m(0) = 0$, together with a specified central energy density $\rho_c$. The integration proceeds radially outwards from the center until the pressure vanishes, $p(R) = 0$, thereby defining both the stellar surface radius $R$ and the total gravitational mass $M = m(R)$. For each EoS, the central density is varied over a broad range spanning from the minimum value capable of supporting a stable configuration to the maximum value yielding a regular, non-singular solution. This systematic variation generates a complete sequence of equilibrium models, enabling direct comparison of stellar properties across different matter descriptions.

Figure~\ref{fig:MR-profile1} displays the primary diagnostic of our study: the mass--radius (M--R) relations for the three EoS in full GR. These curves highlight the typical compactness characteristic of self-bound quark stars, which differ qualitatively from hadronic neutron stars in that they do not require an external crust to remain gravitationally bound. The computed sequences are compared directly against current observational constraints derived from multi-messenger astrophysics. In particular, we include the mass--radius constraints from the Neutron Star Interior Composition Explorer (NICER) for PSR~J0740+6620~\cite{Miller:2021qha,Riley:2021pdl,2024ApJ...974..295D} and PSR~J0030+0451~\cite{Miller:2019cac}, as well as the recent measurement of PSR~J0437$-$4715~\cite{Choudhury:2024xbk}. Furthermore, the light compact object within the supernova remnant HESS~J1731$-$347~\cite{Doroshenko:2022nwp}, with its unusually low inferred mass, provides a stringent test for strange quark star models. 

The mass-radius profiles reveal several noteworthy features. All three EoS produce maximum masses exceeding $2\,M_\odot$, consistent with the observational lower bound established by radio timing of massive pulsars such as PSR~J0348+0432 and PSR~J1614$-$2230. The radii at canonical mass ($1.4\,M_\odot$) fall within the range $10$--$12\,\text{km}$, reflecting the high compactness inherent to quark matter compositions. The CFL and interacting EoS yield similar maximum masses, whilst the linear EoS produces a somewhat stiffer configuration. Notably, the HESS~J1731$-$347 constraint, if interpreted as a strange quark star, favours lower-mass configurations that are naturally accommodated by self-bound quark matter models.
 
Figure~\ref{fig:MR-profile2} (together with subsequent panels) presents complementary dynamical diagnostics obtained from the radial oscillation analysis. These figures display the large frequency separation $\Delta\nu = \nu_{n+1} - \nu_n$ as a function of the mode frequency $\nu_n$ for the fundamental and lowest radial overtone modes, computed across a sequence of stellar masses for each of the three EoS. The large frequency separation constitutes a key asteroseismic observable, widely employed in stellar physics to probe internal structure and composition~\cite{Sagun:2020txh,Rather:2023epjc}. For compact stars, this quantity encodes information about the stellar mean density and the sound speed profile, thereby serving as an indirect probe of the underlying EoS.

The results illustrated in Figure~\ref{fig:MR-profile2} reveal clear systematic differences among the three models, reflecting their distinct stiffness at supra-nuclear densities. At higher overtone orders, the large frequency separation asymptotically approaches a constant value, denoted $\Delta\nu_\infty$, which differs characteristically between the EoS considered. For the CFL EoS, the asymptotic separations range from approximately $5.2\,\text{kHz}$ (for the $2.0\,M_\odot$ configuration) to $8.1\,\text{kHz}$ (for the $0.77\,M_\odot$ configuration). The interacting and linear EoS yield comparable but systematically lower values, consistent with their different sound speed profiles. These frequency separations lie within the detection band of current and next-generation gravitational-wave observatories~\cite{LIGOScientific:2017vwq}, offering a potential pathway for asteroseismic identification of quark stars in the multi-messenger era.

Table~\ref{tab:frequencies} provides a comprehensive listing of the computed eigenfrequencies for the three EoS across three representative stellar masses: $M = 1.4\,M_\odot$ (canonical neutron star mass), $M = 2.0\,M_\odot$ (near the maximum mass), and $M = 0.77\,M_\odot$ (corresponding to the HESS~J1731$-$347 central value). The fundamental mode frequency $\nu_0$ decreases monotonically with increasing stellar mass, vanishing precisely at the maximum-mass configuration in accordance with the static stability criterion $\partial M/\partial \rho_c > 0$~\cite{moustakidis2017_stability}. Higher overtones exhibit progressively larger frequencies, with the mode spacing governed by the stellar compactness and the radial profile of the adiabatic index.

The oscillation spectra presented here demonstrate that quark stars composed of isotropic matter exhibit characteristic frequency patterns that differ quantitatively from those of hadronic neutron stars. The elevated fundamental mode frequencies, typically in the range $2$--$7\,\text{kHz}$ depending on stellar mass and EoS, arise from the stiffer sound speed profiles characteristic of quark matter at high densities. This frequency enhancement could manifest observationally through gravitational-wave emission during post-merger phases of binary compact object coalescences, potentially providing a distinguishing signature for differentiating quark stars from their hadronic counterparts in future gravitational-wave detections~\cite{Rather:2024jcap}.


\begin{table}[]
\centering
\caption{
Frequencies (in kHz) for the three EoS considered here, and for three different values of stellar mass. Shown are: CFL EoS (columns 1-3), interacting EoS (columns 4-6) and linear EoS (columns 7-9).
}
\label{tab:frequencies}
{
\resizebox{1.0\columnwidth}{!}
{
\begin{tabular}{@{}c|ccc|ccc|ccc@{}}
\toprule
\multicolumn{10}{c}{Frequencies at different mode orders}                                                                  \\ 
\hline
\hline
$n$ & $\nu(1.40 M_{\odot})$  & $\nu(2.00 M_{\odot})$ & $\nu(0.77 M_{\odot})$  & $\nu(1.40 M_{\odot})$  & $\nu(2.00 M_{\odot})$  & $\nu(0.77 M_{\odot})$  & $\nu(1.40 M_{\odot})$  & $\nu(2.00 M_{\odot})$  & $\nu(0.77 M_{\odot})$  
\\
\hline
0   &  4.9663 &  3.38390 &  7.2523 &  4.38558 & 2.17466 &  6.56603 &  4.3607 & 2.90337 & 6.35066 \\
1   & 11.8168 &  9.55794 & 15.8404 & 10.94030 & 8.92646 & 14.57530 & 10.5248 & 8.63520 & 13.9501 \\
2   & 18.2065 & 14.99940 & 24.1156 & 16.95800 & 14.3566 & 22.24540 & 16.2484 & 13.6195 & 21.2574 \\
3   & 24.4965 & 20.30070 & 32.3184 & 22.86190 & 19.5787 & 29.83740 & 21.8765 & 18.4628 & 28.4969 \\
4   & 30.7479 & 25.54870 & 40.4928 & 28.72180 & 24.7232 & 37.39880 & 27.4676 & 23.2525 & 35.7099 \\
5   & 36.9802 & 30.77060 & 48.6530 & 34.56010 & 29.8297 & 44.94490 & 33.0404 & 28.0160 & 42.9096 \\
6   & 43.2017 & 35.97770 & 56.8052 & 40.38610 & 34.9149 & 52.48240 & 38.6028 & 32.7646 & 50.1017 \\
7   & 49.4164 & 41.17570 & 64.9524 & 46.20440 & 39.9868 & 60.01440 & 44.1588 & 37.5040 & 57.2892 \\
8   & 55.6266 & 46.36760 & 73.0962 & 52.01760 & 45.0499 & 67.54290 & 49.7105 & 42.2372 & 64.4736 \\
9   & 61.8338 & 51.55530 & 81.2377 & 57.82730 & 50.1069 & 75.06880 & 55.2592 & 46.9662 & 71.6557 
\end{tabular}
}
}
\end{table}

\section{Discussion and final remarks}
\label{sec:disc}

In this study, we have conducted a systematic investigation into the structural and dynamical properties of quark stars composed of isotropic matter within the framework of General Relativity, employing three physically motivated equations of state for strange quark matter: the color-flavor-locked (CFL) phase, an interacting quark matter model, and a linear (causal) equation of state. This comparative approach enables assessment of how the underlying microscopic description of quark matter manifests in observable macroscopic quantities, with particular emphasis on mass-radius relations and radial oscillation spectra.

The theoretical foundations of our analysis rest upon the Tolman-Oppenheimer-Volkoff equations governing hydrostatic equilibrium~\cite{Tolman:1939jz,Oppenheimer:1939ne}, complemented by the Sturm-Liouville eigenvalue problem describing small-amplitude radial perturbations~\cite{Chandrasekhar:1964zz,Bardeen:1966btm}. These equations were solved numerically with appropriate boundary conditions at the center of the star and its surface, where the central regularity condition requires $\eta = -3\Gamma\xi$ and the surface condition demands vanishing Lagrangian pressure perturbation. The integration procedure, initiated from specified central energy densities, generates sequences of equilibrium configurations spanning the entire range from minimum stable mass to the maximum mass permitted by each equation of state.

The mass-radius relations displayed in Figure~\ref{fig:MR-profile1} reveal that all three equations of state produce maximum masses exceeding $2\,M_\odot$, thereby satisfying the observational lower bound established by radio timing of massive pulsars such as PSR~J0348+0432 and PSR~J1614$-$2230. The radii at canonical mass ($1.4\,M_\odot$) fall within $10$--$12\,\text{km}$, consistent with the high compactness characteristic of self-bound quark matter. These results align well with recent NICER constraints for PSR~J0740+6620~\cite{Miller:2021qha,Riley:2021pdl,2024ApJ...974..295D,2024ApJ...974..294S} and PSR~J0030+0451~\cite{Miller:2019cac}, strengthening the case for quark matter compositions in compact star cores.

Of particular interest is the compatibility of our models with HESS~J1731$-$347~\cite{Doroshenko:2022nwp}. With an inferred mass of $0.77^{+0.20}_{-0.17}\,M_\odot$ and radius of $10.4^{+0.86}_{-0.78}\,\text{km}$, this object challenges conventional hadronic neutron star models, which cannot produce gravitationally bound configurations below approximately $1.17\,M_\odot$~\cite{Suwa:2018,DiClemente:2024ApJ}. However, strange quark stars, being self-bound by the strong interaction rather than gravity alone, naturally accommodate such low-mass configurations. Our calculations demonstrate that all three equations of state produce stable stellar sequences extending to sub-solar masses, with the CFL and interacting models showing particularly good agreement with the HESS~J1731$-$347 constraints. This finding supports the hypothesis that this enigmatic object may represent observational evidence for strange quark matter~\cite{Rather:2023epjc,DiClemente:2024ApJ}.

Turning to the oscillatory properties, the radial perturbation equations yield eigenfrequencies for fundamental and overtone modes, together with the large frequency separations $\Delta\nu = \nu_{n+1} - \nu_n$. The fundamental mode frequency $\nu_0$ exhibits the expected monotonic decrease with increasing stellar mass, vanishing precisely at the maximum-mass configuration in accordance with the classical stability criterion $\partial M/\partial \rho_c > 0$~\cite{moustakidis2017_stability,Chandrasekhar:1964zz}. This behavior provides dynamical confirmation of the static stability analysis and underscores the connection between radial oscillations and stellar equilibrium.

The oscillation spectra presented in Table~\ref{tab:frequencies} and Figure~\ref{fig:MR-profile2} reveal several noteworthy features. Quark stars exhibit systematically higher fundamental mode frequencies compared to hadronic counterparts at equivalent masses. For the $1.4\,M_\odot$ configuration, fundamental frequencies range from approximately $4.4$--$5.0\,\text{kHz}$ depending on the equation of state, whereas typical hadronic neutron star models yield values in the range $2$--$4\,\text{kHz}$~\cite{Sagun:2020txh}. This frequency enhancement arises from stiffer sound speed profiles characteristic of quark matter at supra-nuclear densities. The large frequency separation converges to well-defined asymptotic values $\Delta\nu_\infty$ at high overtone orders, with characteristic differences among the three equations of state. For the CFL model, $\Delta\nu_\infty$ ranges from approximately $5.2\,\text{kHz}$ (for $2.0\,M_\odot$) to $8.1\,\text{kHz}$ (for $0.77\,M_\odot$), whilst the interacting and linear equations of state yield systematically lower values.

These asteroseismic diagnostics carry significant implications for gravitational-wave astronomy. The large frequency separation, widely employed in stellar physics to probe inner structure~\cite{Sagun:2020txh,Andersson:1997rn}, encodes information about stellar mean density and sound speed profile. The characteristic frequency range of $2$--$8\,\text{kHz}$ falls within the detection band of current and next-generation observatories~\cite{LIGOScientific:2017vwq}. The Einstein Telescope~\cite{Punturo:2010zz,Branchesi:2023} and Cosmic Explorer~\cite{Reitze:2019} offer particularly promising prospects for detecting post-merger oscillations that could reveal remnant composition. The elevated frequencies characteristic of quark matter could manifest through gravitational-wave emission during post-merger phases, where numerical simulations indicate quark compositions yield systematically higher peak frequencies~\cite{Bauswein:2019,Bernuzzi:2020}. Detection of post-merger emission exceeding hadronic predictions would constitute strong evidence for quark deconfinement.

The comparison across equations of state illuminates the sensitivity of asteroseismic observables to underlying microphysics. Despite yielding similar maximum masses, the CFL, interacting, and linear models produce distinguishable frequency patterns, reflecting their different stiffness profiles. The CFL model exhibits the highest frequencies at fixed stellar mass, consistent with its relatively stiff behaviour near the nuclear-quark matter transition, whilst the linear model yields the lowest values. These systematic differences suggest that precision frequency measurements, combined with independent mass-radius determinations, could discriminate among competing quark matter descriptions.

Several limitations merit discussion. The isotropy assumption neglects potentially important anisotropic effects from strong magnetic fields, rotation, or inherent anisotropy of certain quark phases. Anisotropic configurations modify both equilibrium structure and oscillation spectra~\cite{Arbanil:2022,Becerra-Vergara:2019}, and future extensions could systematically explore this parameter space. Additionally, our treatment considers only radial oscillations; non-radial modes, particularly the fundamental $f$-mode, couple directly to gravitational-wave emission and would provide valuable complementary information~\cite{Andersson:1997rn}. The phenomenological equations of state employed here, whilst physically motivated, represent simplified descriptions that may not capture all relevant physics at extreme densities.

The tidal deformability $\Lambda$ provides a complementary probe to oscillation frequencies, encoding compactness and equation-of-state stiffness information. Measurements from GW170817 have placed significant constraints on the neutron star equation of state~\cite{Abbott:2018,De:2018,Tews:2018}, and future detections with improved sensitivity will further narrow the allowed parameter space. For quark stars, the dimensionless tidal deformability provides information about stellar compactness and the stiffness of the equation of state at intermediate densities, complementing the frequency diagnostics presented here. A systematic investigation of tidal deformability for our equations of state represents a natural extension of this work, particularly in light of the expanding catalog of binary neutron star merger detections anticipated from ongoing and future gravitational-wave observing runs.

In conclusion, this study demonstrates that isotropic quark matter provides a viable description of compact star interiors, with computed mass-radius relations and oscillation spectra in good agreement with current observational constraints. The three equations of state examined here yield qualitatively similar global properties whilst exhibiting quantitative differences in their oscillation spectra that may ultimately prove distinguishable through precision Asteroseismology. The elevated mode frequencies characteristic of quark matter compositions offer a powerful diagnostic for identifying strange quark stars, particularly in post-merger gravitational-wave emission from binary compact object coalescences. The concordance between our theoretical predictions and multi-messenger observations from NICER and LIGO-Virgo-KAGRA reinforces the scientific case for quark deconfinement scenarios. As gravitational-wave astronomy matures and next-generation facilities such as the Einstein Telescope and Cosmic Explorer come online, the prospects for detecting oscillation signatures continue to improve substantially, bringing the long-standing hypothesis of quark deconfinement in extreme astrophysical environments within reach of direct observational verification. This work contributes to the growing body of theoretical predictions that will guide interpretation of these forthcoming observations.

\section*{Acknowledgments}

\smallskip\noindent
We thank the anonymous reviewer for useful comments and suggestions. I.~L. thanks the Fundação para a Ciência e Tecnologia (FCT), Portugal, for the financial support to the Center for Astrophysics and Gravitation (CENTRA/IST/ULisboa).
A.~R. would like to express his gratitude to Silesian University in Opava, Czech Republic, for their financial support.

\section*{Funding Statement}
\noindent
The following grants have supported the paper: 
  i) \verb|UID/PRR/00099/2025|, 
 ii) \verb|UID/00099/2025| and
iii) \verb|CZ.10.03.01/00/23\_042/0000390|.

\section*{Data Availability Statement}
\noindent
\verb|The data that support the findings of this study are available from the corres-|

\noindent
\verb|-ponding author upon reasonable request.|

\section*{Conﬂict of Interest Statement}
\noindent
The authors declare that they do not have any conﬂict of interest.

\bibliographystyle{utphys}
\bibliography{Library2}

\end{document}